\documentclass[12pt]{article}

\usepackage[table]{xcolor} 

\usepackage[margin=1in]{geometry} 
\usepackage{setspace} 
\usepackage{xr}
\usepackage{subcaption}
\usepackage[labelsep=quad,skip=-2pt]{caption}
\usepackage[utf8]{inputenc} 
\usepackage[T1]{fontenc}    
\usepackage{hyperref}       
\usepackage{url}            
\usepackage{booktabs}       
\usepackage{amsfonts}       
\usepackage{nicefrac}       
\usepackage{microtype}      
\usepackage{lipsum}
\usepackage{graphicx,psfrag,epsf}
\graphicspath{ {./Fig/} }
\usepackage{enumerate}
\usepackage[justification=centering]{caption}
\usepackage{tikz}
\usetikzlibrary{matrix, positioning, arrows.meta}

\usepackage{float}

\usepackage{bm}
\usepackage{amsmath}
\usepackage{amsfonts}
\usepackage{amssymb}

\usepackage{amsthm}
\usepackage{amssymb}

\newtheorem{Theorem}{Theorem}

\newtheorem{Assumption}{Assumption}

\DeclareMathOperator{\logit}{logit}

\usepackage{color}

\usepackage[authoryear]{natbib}
\usepackage{comment}

\usepackage[ruled,vlined]{algorithm2e}
\usepackage{authblk} 

\title{A sensitivity analysis approach to principal stratification with a continuous longitudinal intermediate outcome: Applications to a cohort stepped wedge trial}

\author[1]{Lei Yang}
\author[1]{Michael J. Daniels\thanks{Corresponding author. Email: \texttt{daniels@ufl.edu}}}
\author[2]{Fan Li}

\affil[1]{Department of Statistics, University of Florida, Gainesville, FL, USA}
\affil[2]{Department of Biostatistics, Yale School of Public Health, New Haven, CT, USA}

\date{}

\begin{document}
\maketitle

\setlength{\abovedisplayskip}{5pt}
\setlength{\belowdisplayskip}{5pt}
\setlength{\abovedisplayshortskip}{5pt}
\setlength{\belowdisplayshortskip}{5pt}

\begin{abstract}
Causal inference in the presence of intermediate variables is a challenging problem in many applications. Principal stratification (PS) provides a framework to estimate principal causal effects (PCE) in such settings. However, existing PS methods primarily focus on settings with binary intermediate variables. We propose a novel approach to estimate PCE with continuous intermediate variables in the context of stepped wedge cluster randomized trials (SW-CRTs). Our method leverages the time-varying treatment assignment in SW-CRTs to calibrate sensitivity parameters and identify the PCE under realistic assumptions. We demonstrate the application of our approach using data from a cohort SW-CRT evaluating the effect of a crowdsourcing intervention on HIV testing uptake among men who have sex with men in China, with social norms as a continuous intermediate variable. The proposed methodology expands the scope of PS to accommodate continuous variables and provides a practical tool for causal inference in SW-CRTs.

\end{abstract}


\section{Introduction and literature review}\label{sec:introduction}

In many scientific applications, it is of interest to investigate the causal pathway underlying the total treatment effect when an intermediate variable is present. Several different types of intermediate variables have been studied in the prior literature, including but not limited to treatment compliance \citep{ANGRIST1996,ROY2008,JIN2008}, death as a terminal event \citep{DAI2012, XU2022} and secondary outcomes \citep{KIM2019a}.

There exist several frameworks that can address intermediate variables. Causal mediation analysis aims to decompose the total effect into natural direct and indirect effects by exploring the causal pathway under an intervention on the intermediate variable \citep{IMAI2010}. An alternative approach, principal stratification \citep[PS;][]{FRANGAKIS2002}, focuses on characterizing treatment effect heterogeneity within strata defined by the potential outcomes of the intermediate variable. {Rather than focusing on effect decomposition, PS uses the potential outcomes of the intermediate variable under different treatment arms to define subgroups (principal strata). The principal causal effects (PCEs) are the comparison of the potential outcomes within these subgroups. It is important to note that PCEs, such as the associative effect, generally encompass both direct pathways and pathways through the mediator, rather than isolating the indirect effect itself (\cite{VANDERWEELE2008}).}

Under the latter framework, much of the prior literature focused on point and interval identification of the PCEs with a binary intermediate variable. A popular method for identification is based on an instrumental variable (\cite{ANGRIST1996}), typically under the monotonicity and the exclusion restriction assumptions. \cite{ROY2008} introduced a weaker version of the monotonicity assumption. {\cite{DING2017,jiang2022multiply} and \citet{tong2025semiparametric} replaced the exclusion restriction assumption with the principal ignorability assumption, and developed the principal score weighting and multiply robust estimators to point identify the PCEs. However, principal ignorability requires a rich set of baseline covariates to adequately deconfound the relationship between the intermediate variable and the outcome, which is not always feasible in practice. Furthermore, our motivating closed-cohort stepped wedge trial only includes limited amount of baseline information, rendering the plausibility of the principal ignorability assumption questionable. Furthermore, our motivating study features a continuous intermediate outcome and clustered longitudinal data, for which the existing methods under principal ignorability may not be directly applicable.} To relax the untestable structural assumptions, others have also used a parametric mixture approach to empirically identify the PCEs with independent and clustered data (e.g., \cite{IMBENS1997, tong2023bayesian}).

Continuous intermediate variables introduce additional challenges for principal stratification analysis, compared to binary or categorical variables. Most of the existing methods either dichotomized the intermediate variable (\cite{BACCINI2017}), or assumed a fully parametric model for the joint distribution of the potential intermediate variables (\cite{MAGNUSSON2019}). However, as discussed in \cite{SCHWARTZ2011}, the former is subject to information loss and arbitrary choice of the cutoff point and the latter is often inadequate to represent complex distributional and clustering features. In contrast, \cite{SCHWARTZ2011} treated principal stratification as an incomplete data problem, and use a Dirichlet process mixture to address latent clustering features. \cite{KIM2019a} studied multiple continuous intermediate variables and propose a Gaussian copula assumption and an additional homogeneity assumption to identify the PCE. \cite{ANTONELLI2023} extended the principal stratification framework to studies with continuous treatments and continuous intermediate variables.

While there has been extensive development of principal stratification methods with cross-sectional data, relatively fewer efforts have focused on principal stratification with longitudinal data, with the following few exceptions. \cite{YAU2001} studied the causal effects in the presence of treatment noncompliance in a longitudinal setting. Their framework focused on a time-fixed treatment and assumed away defiers and always-takers. \cite{FRANGAKIS2004} considered studies with time-varying treatment and compliance. However, their approach did not account for possible correlations among repeated measurements taken on the same unit.
Their identification result depends on two strong assumptions, multilevel monotonicity and a compound exclusion restriction, which are generalization of standard monotonicity and exclusion restriction assumptions to accommodate a time-varying treatment. \cite{LIN2008, LIN2009} proposed a hierarchical latent class structure that consists of time-varying compliance nested in classes of longitudinal compliance trends that are time-invariant in a parametric setting. \cite{DAI2012} considered a partially Hidden Markov model on compliance behavior with a time-to-event endpoint. Despite their focus on the longitudinal data structure, these prior efforts have been restricted to a binary intermediate variable---typically noncompliance, and have not been expanded to address a continuous intermediate variable. 

\section{Motivating data example and objective}\label{sec:motivating}

Our study is motivated by a closed-cohort stepped wedge cluster randomized trial (SW-CRT) in the presence of a continuous intermediate variable. SW-CRTs are a recent variant of cluster randomized designs that are increasingly common for evaluating healthcare interventions (\cite{nevins2024adherence}). In a SW-CRT, clusters are randomized to different time points corresponding to when the intervention starts. All clusters start in the control condition and eventually receive the treatment. These features allow each cluster to serve as its own control and can facilitate cluster recruitment and stakeholder engagement especially when the intervention is perceived to be beneficial. \citet{TANG2018} reported a completed SW-CRT which evaluated the impact of a newly-developed crowdsourcing HIV intervention on HIV testing uptake among men who have sex with men in eight Chinese cities, from August 2016 to August 2017. The crowdsourcing intervention included a multimedia HIV testing campaign, an online HIV testing service, and local testing promotion campaigns tailored for men who have sex with men. The study employed a closed-cohort design and recruited men who have sex with men prior to randomization of cities. The intervention was initiated for pairs of cities at 3-month intervals, and each pair of cities received the intervention for 3 consecutive months. In total, the study collected data at baseline followed by four time points over 12 months, and enrolled a total of 1,381 participants as a closed cohort.

The primary outcome of this study was the proportion of participants who tested for HIV over the previous 3 months. Table~\ref{tab:table_YM}(a) shows the HIV testing proportion in different time periods. An important secondary outcome was the sensitivity to HIV testing social norms, which was measured by six items asking participants about their perceived social norm regarding HIV testing. {While the cluster-period means in Table \ref{tab:table_YM}(b) suggest a modest average impact of the intervention on social norms, this does not necessarily imply an absence of intervention effect. First, observed averages may conflate secular trends with intervention effects in a stepped wedge trial, and opposing directions may mask true changes due to the crowdsourcing intervention. Second, the absence of a strong mean effect does not rule out important heterogeneity across participants, as some may experience increases in perceived norms while others experience decreases. The principal stratification framework is well-suited to investigate potential treatment effect heterogeneity within these different response patterns. Finally, the HIV testing social norms have been studied as an intermediate outcome (or mediator) and proved to have a significant relationship with the uptake of HIV testing in prior observational studies \citep{PERKINS2018, ZHAO2018}, providing some basis for considering social norm as an intermediate outcome of interest in this trial.}
In the context of the SW-CRT reported by \citet{TANG2018}, our goal is to investigate whether the causal effects of crowdsourcing intervention on the HIV testing uptake differ across different strata defined by social norm, under a principal stratification framework.

\begin{table}[ht!]
\caption{Summary statistics on HIV testing proportion (\%) and social norms by city and time periods. Shaded cells represent intervention cluster-periods and white cells represent control cluster-periods.} 
\vspace{1em}
    \centering
    \begin{subtable}{\textwidth}
        \centering
        \caption{HIV testing proportion (\%); In the baseline period ($t=1$), the HIV testing proportion is zero. Data format: \% (Count/N).}
        \resizebox{\textwidth}{!}{
        \begin{tabular}{lccccc}
            \toprule
            City & $t=1$ (N) & $t=2$ & $t=3$ & $t=4$ & $t=5$ \\
            \midrule
            1. Guangzhou & 203 & 16.9 (26/154) & \cellcolor{gray!50}37.6 (56/149) & \cellcolor{gray!50}23.7 (37/156) & \cellcolor{gray!50}35.1 (52/148) \\
            2. Yantai    & 180 & 21.6 (30/139) & \cellcolor{gray!50}32.8 (43/131) & \cellcolor{gray!50}27.5 (33/120) & \cellcolor{gray!50}28.4 (36/127) \\
            3. Jiangmen  & 139 & 17.5 (21/120) & 33.6 (40/119) & \cellcolor{gray!50}30.8 (33/107) & \cellcolor{gray!50}37.3 (38/102) \\
            4. Jinan     & 189 & 21.4 (34/159) & 31.9 (45/141) & \cellcolor{gray!50}27.7 (36/130) & \cellcolor{gray!50}35.9 (47/131) \\
            5. Zhuhai    & 134 & 20.7 (23/111) & 27.4 (31/113) & 44.6 (49/110) & \cellcolor{gray!50}40.6 (43/106) \\
            6. Qingdao   & 182 & 19.2 (28/146) & 21.1 (30/142) & 54.1 (73/135) & \cellcolor{gray!50}38.5 (50/130) \\
            7. Shenzhen  & 203 & 20.4 (33/162) & 25.5 (41/161) & 28.1 (45/160) & \cellcolor{gray!50}50.7 (75/148) \\
            8. Jining    & 151 & 22.5 (29/129) & 31.8 (42/132) & 30.2 (38/126) & \cellcolor{gray!50}46.1 (53/115) \\
            \bottomrule
        \end{tabular}
        }
    \end{subtable}
    
    \vspace{0.5 em}
    
    \begin{subtable}{\textwidth}
        \centering
        \caption{HIV testing social norms summarized by cluster-period means. Data format: Mean (N).}
        \resizebox{\textwidth}{!}{
        \begin{tabular}{lccccc}
            \toprule
            City & $t=1$ & $t=2$ & $t=3$ & $t=4$ & $t=5$ \\
            \midrule
            1. Guangzhou & 14.6 (203) & 14.3 (154) & \cellcolor{gray!50}14.9 (149) & \cellcolor{gray!50}14.8 (156) & \cellcolor{gray!50}14.5 (148) \\
            2. Yantai    & 14.9 (180) & 14.7 (139) & \cellcolor{gray!50}14.5 (131) & \cellcolor{gray!50}14.7 (120) & \cellcolor{gray!50}14.7 (127) \\
            3. Jiangmen  & 15.0 (139) & 14.6 (120) & 15.1 (119) & \cellcolor{gray!50}14.5 (107) & \cellcolor{gray!50}14.6 (102) \\
            4. Jinan     & 14.6 (189) & 14.4 (159) & 14.3 (141) & \cellcolor{gray!50}14.2 (130) & \cellcolor{gray!50}14.3 (131) \\
            5. Zhuhai    & 14.7 (134) & 14.7 (111) & 15.0 (113) & 14.6 (110) & \cellcolor{gray!50}15.1 (106) \\
            6. Qingdao   & 14.9 (182) & 14.8 (146) & 14.7 (142) & 14.9 (135) & \cellcolor{gray!50}14.8 (130) \\
            7. Shenzhen  & 14.6 (203) & 14.5 (162) & 14.8 (161) & 14.6 (160) & \cellcolor{gray!50}14.9 (148) \\
            8. Jining    & 14.6 (151) & 14.3 (129) & 14.6 (132) & 14.2 (126) & \cellcolor{gray!50}14.5 (115) \\
            \bottomrule
        \end{tabular}
        }
    \end{subtable}
    \label{tab:table_YM} 
\end{table}

To provide insights into the role of social norm in the crowdsourcing HIV intervention study, we pursue the potential outcomes framework, {and propose an approach to define principal causal effect estimands in the context of closed-cohort SW-CRTs with a continuous intermediate outcome. Our framework accommodates arbitrary treatment contrasts, allowing the assessment of both short-term and long-term principal causal effects (PCEs).}

We then propose new structural assumptions to achieve point identification of the principal causal effect estimands under a sensitivity framework. These include a copula assumption that addresses the joint distribution of the potential intermediate variables under treatment and control conditions, and a marginal structural assumption that addresses the relationship between potential outcome and potential intermediate variables. To implement our procedure, we exploit a unique feature of SW-CRTs that the intermediate variable and outcome are observable in both treatment arms at different time points, which provides useful information to calibrate values for sensitivity parameters. 
This calibration approach offers a practical advantage over methods relying on untestable structural assumptions, such as principal ignorability (\cite{DING2017}), by grounding the sensitivity analysis in observed data.

We then consider random-effects models that incorporate duration-specific treatment effects \citep{LI2021,WANG2024} and pursue a Bayesian framework for inference. The Bayesian inferential framework is attractive as it can accommodate monotone missing data (under an additional ignorability assumption), which is present in the crowdsourcing HIV intervention study. 
Different from the existing literature discussed in Section \ref{sec:introduction}, our development represents the first effort to simultaneously addresses the longitudinal data structure (arising from the closed-cohort stepped wedge design), accommodate a continuous intermediate outcome, and provide a generalized framework for assessing both short- and long-term principal causal effects.

The remainder of this paper is organized as follows. Section \ref{sec:methods} introduces our proposed methodology, including notation specific for closed-cohort SW-CRTs, proposed causal estimands, structural assumptions, and new identification result. Section \ref{sec:model} discusses the observed data modeling approach and methods to calibrate sensitivity parameters using observed data. In Section \ref{sec:analysis}, we apply our proposed methodology to analyze the crowdsourcing HIV intervention study to provide additional insights into the role of social norm in explaining the causal effects on HIV testing uptake. Finally, Section \ref{sec:discussion} provides a discussion of the implications of our findings and potential areas for future research.

\section{A sensitivity analysis framework for principal stratification with a longitudinal continuous intermediate variable}\label{sec:methods}

\subsection{Notation for SW-CRTs}

We consider a closed-cohort SW-CRT where all participating individuals are identified prior to cluster randomization. We first discuss our proposed methods in the absence of attrition; extensions to address monotone missing data will be discussed in Section \ref{sec:posterior} in the context of the crowdsourcing HIV intervention study. We use $i=1,\ldots,N_j$ to denote each individual, $j=1,\ldots,J$ to denote cluster and $t=1,\ldots,T$ to denote the discrete time period. As we consider a closed-cohort design, 
individuals are nested in clusters, which are cross-classified by periods. Let $Z_{jt}$ be the indicator of intervention status of cluster $j$ at time $t$. Due to the staggered rollout design feature, $Z_{jt} \geq Z_{j,t-1}$ for any $j$ and $t$. Furthermore, we let $M_{ijt}$ and $Y_{ijt}$ denote the intermediate variable and the outcome for individual $i$ in cluster $j$ measured during period $t$, and we assume that the cluster-level intervention $Z_{jt}$ happens prior to $M_{ijt}$, which also temporally precedes $Y_{ijt}$ and hence is an intermediate outcome. {Additionally, we define $\bm X_{ij}$ as a set of time-invariant baseline covariates for each individual, and $\bm C_{j}$ as a set of time-invariant, baseline covariates for each cluster. We use the overbar notation to denote history of variables. According, $\overline{\bm M}_{ijt} = (M_{ij1}, M_{ij2}, \dots, M_{ijt})$, $\overline{\bm Y}_{ijt} = (Y_{ij1}, Y_{ij2}, \dots, Y_{ijt})$, and $\overline{\bm Z}_{jt} = (Z_{j1}, Z_{j2}, \dots, Z_{ijt})$ represent the history of intermediate variables up to time $t$, the history of outcomes variables up to time $t$, and the history of treatment variables up to time $t$, respectively; $\overline{\bm M}_{ijT}$, $\overline{\bm Y}_{ijT}$, and $\overline{\bm Z}_{jT}$ are thus the full histories over the entire study period.} 

\subsection{Standard assumptions for SW-CRTs}

We use a potential outcomes framework \citep{RUBIN1974} to clarify several standard assumptions for closed-cohort SW-CRTs. Let $\overline{\bm{Z}} = (\overline{\bm{Z}}_{1T}, \overline{\bm{Z}}_{2T}, \dots, \overline{\bm{Z}}_{JT})$ be a $T\times J$ matrix of treatment assignments, where $\overline{\bm{Z}}_{jT} \in \mathcal{\bm Z}$ with $\mathcal{\bm Z}$ being its support. Let $\overline{\bm{M}}_{ijT}(\overline{\bm{Z}})$ be the vector of potential outcomes of the intermediate variable for individual $i$ in cluster $j$ given treatment $\overline{\bm{Z}}$, and similarly for $\overline{\bm{Y}}_{ijT}(\overline{\bm{Z}})$. We first make the following assumption of no interference among clusters \citep{CHEN2024}.
\begin{Assumption}[Cluster-level SUTVA]
    \label{assumption:sutva}
    {For any two matrices of treatment assignments, $\overline{\bm{Z}}, \overline{\bm{Z}}^*$, if $\overline{\bm{Z}}_{jT} = \overline{\bm{Z}}_{jT}^*$, then $\overline{\bm{M}}_{ijT}(\overline{\bm{Z}}) = \overline{\bm{M}}_{ijT}(\overline{\bm{Z}}^*)$, and $\overline{\bm{Y}}_{ijT}(\overline{\bm{Z}}) = \overline{\bm{Y}}_{ijT}(\overline{\bm{Z}}^*)$.}
\end{Assumption}

{Under Assumption \ref{assumption:sutva}, the potential outcomes for individual $i$ in cluster $j$ only depend on the treatment sequence assigned to their own cluster. We can therefore simplify the notation as $\overline{\bm{Y}}_{ijT}(\overline{\bm z}_{jT})$ and $\overline{\bm{M}}_{ijT}(\overline{\bm z}_{jT})$. We next introduce the following sampling and randomization assumptions.}
{\begin{Assumption}[Super-population sampling]
    \label{assumption:sampling}
    Denote $N_j$ as the size of the closed cohort in cluster $j$, and $\overline{\bm{W}}_j = \{\overline{\bm{Y}}_{ijT}(\overline{\bm z}_{T}), \overline{\bm{M}}_{ijT}(\overline{\bm z}_{T}), \bm X_{ij}, \bm C_j, N_j : i=1,\dots,N_j, \overline{\bm z}_{T}\in\mathcal{\bm Z}\}$ as the complete data vector for cluster $j$. Assume $\{\overline{\bm{W}}_1, \dots, \overline{\bm{W}}_J\}$ are independent and identically distributed draws from a population distribution with finite second moments. Furthermore, the data vectors $\{\overline{\bm{Y}}_{ijT}(\overline{\bm z}_{T}), \overline{\bm{M}}_{ijT}(\overline{\bm z}_{T}), \bm X_{ij} : \overline{\bm z}_{jT}\in\mathcal{\bm Z}\}$ for $i=1,\dots,N_j$ are assumed identically distributed given cluster-level information $\bm C_j$ and $N_j$.
\end{Assumption}
{
\begin{Assumption}[Staggered randomization]
    \label{assumption:randomization}
    For any possible treatment sequence $\overline{\bm z}_{T}\in\mathcal{\bm Z}$,
    \begin{align*}
        \overline{\bm{Z}}_{jT} \perp \{\overline{\bm{Y}}_{ijT}(\overline{\bm z}_{T}), \overline{\bm{M}}_{ijT}(\overline{\bm z}_{T}), \bm X_{ij} : i=1,\dots,N_j, \overline{\bm z}_{T}\in\mathcal{\bm Z}\} \mid \{\bm C_j, N_j\}
    \end{align*}
\end{Assumption}
}
}

{Assumption \ref{assumption:sampling} extends Assumption A1 in \citet{WANG2024} to accommodate an additional intermediate variable. This condition is primarily technical; it ensures that causal estimands are well-defined and provides a framework for statistical inference. Assumption \ref{assumption:randomization} states that the assignment of treatment sequences is random, conditional on all available baseline cluster-level information. This assumption holds in our motivating application where cities were randomized to sequences within provinces \citep{TANG2018}. 
}
{
\begin{Assumption}[No anticipation]
    \label{assumption:no_anticipation}
    For any two treatment sequences $\overline{\bm z}_{T}, \overline{\bm z}_{T}^*$, let $\overline{\bm z}_{t}$ and $\overline{\bm z}_{t}^*$ be the sub-vectors containing their first $t$ elements, with $t=1,\ldots,T$. If $\overline{\bm z}_{t} = \overline{\bm z}_{t}^*$, then $Y_{ijt}(\overline{\bm z}_{T}) = Y_{ijt}(\overline{\bm z}_{T}^*)$ and $M_{ijt}(\overline{\bm z}_{T}) = M_{ijt}(\overline{\bm z}_{T}^*)$.
\end{Assumption}
}

{Assumption \ref{assumption:no_anticipation} rules out the possibility that potential outcomes at time $t$ are affected by treatment assignments in future periods (from period $t+1$ to $T$). This assumption is plausible in our motivating application for two reasons. First, the rollout of the intervention was randomized at the city level and the implementation schedule was not known to participants in advance. Second, because the campaign materials were released only at the start of the intervention period in each city, there was limited opportunity for individuals to alter HIV testing behaviors prior to campaign launch. These features suggest that anticipation effects are unlikely in our data example. Conceptually, it is important to distinguish Assumption \ref{assumption:no_anticipation} from the absence of delayed or cumulative effects. As reflected in our causal estimands in Section \ref{sec:PO}, delayed or cumulative responses to earlier assignments are fully compatible with Assumption \ref{assumption:no_anticipation} and are explicitly modeled through the duration-specific effects in Section \ref{sec:model}. In other words, ``no anticipation'' refers to the exclusion of forward-looking behavioral or physiological responses to future assignments, rather than to the absence of temporal persistence of prior assignments \citep{wang2025anticipation}.} 

\subsection{Principal stratification}
\label{sec:PO}

Principal stratification compares the potential outcomes for subgroups defined by the potential values of the intermediate variable \citep{FRANGAKIS2002}. In a SW-CRT, the cluster-level treatment is a time-varying exogenous variable. 
{To formally define our causal estimands, we first introduce some notation for treatment histories. Let $\overline{\bm e}_{l:t} = (0, \dots, 0, \underbrace{1, \dots, 1}_{\text{$l$th to $t$th element}})$ represent a treatment vector of length $t$ where an intervention begins at period $l$ (for $1 \le l \le t$) and continues through period $t$. A vector of all zeros, $\overline{\bm 0}_t$, represents the control condition up to time $t$. We use this notation to define our causal estimands 
under any two treatment histories, $\overline{\bm z}_t$ and $\overline{\bm z}_t^*$. 
}

{The primary estimands of interest are defined within principal strata formed by the potential outcomes of the intermediate variable under different treatment histories. For any comparison between two treatment histories $\overline{\bm z}_t$ and $\overline{\bm z}_t^*$, we can define period-specific \textit{associative} and \textit{dissociative} effects as follows \citep{VANDERWEELE2008}:
\begin{align*}
    \text{Dissociative effect in period $t$: } 
        & E\left[Y_{ijt}(\overline{\bm z}_t) - Y_{ijt}(\overline{\bm z}_t^*) |
            M_{ijt}(\overline{\bm z}_t) = M_{ijt}(\overline{\bm z}_t^*)\right] \\
    \text{Associative effect in period $t$: } 
        & E\left[Y_{ijt}(\overline{\bm z}_t) - Y_{ijt}(\overline{\bm z}_t^*) |
            M_{ijt}(\overline{\bm z}_t) \neq M_{ijt}(\overline{\bm z}_t^*)\right].
\end{align*}}{In Supplementary Section A, we show that, under the super-population sampling Assumption \ref{assumption:sampling}, these estimands align with the concept of \emph{cluster-average treatment effect estimands} defined in \citet{kahan2024demystifying} and \citet{WANG2024}. }

When the intermediate variable is continuous, the probability of the event $M_{ijt}(\overline{\bm z}_t) = M_{ijt}(\overline{\bm z}_t^*)$ is zero. To accommodate continuous intermediate variables, we therefore define our principal causal effect estimands based on whether the change in the intermediate variable falls within an `indifference' interval $\mathcal{I}$ \citep{ZIGLER2012, KIM2019a}. The generalized PCE in period $t$ comparing treatment histories $\overline{\bm z}_t$ and $\overline{\bm z}_t^*$ for a subpopulation defined by $\mathcal{I}$ is:
\begin{align}
    \label{eq:estimand}
    \text{PCE}_t(\overline{\bm z}_t, \overline{\bm z}_t^*; \mathcal{I}) = 
        E\left[Y_{ijt}(\overline{\bm z}_t) - Y_{ijt}(\overline{\bm z}_t^*) |
            M_{ijt}(\overline{\bm z}_t) - M_{ijt}(\overline{\bm z}_t^*) \in \mathcal{I}\right].
\end{align}
{This general definition encompasses both generalized dissociative and associative effects. For instance, the generalized dissociative effect can be represented by setting $\mathcal{I} = (-\delta, \delta)$ for a small $\delta > 0$. On the other hand, the generalized associative effects are captured when $\mathcal{I}$ includes a meaningful change in the intermediate variable (e.g., $(-\infty, -\delta)$ or $(\delta, \infty)$).}

\begin{table}[h]
\caption{Summary of key notation and estimands definition.}
\vspace{1em}
\centering
\begin{tabular}{ll}
\hline
Notation & Description \\
\hline
$\overline{\bm z}_{t}$ & Treatment history up to time t. \\
$\overline{\bm 0}_{t}$ & Control condition up to time t (vector of zeros). \\
$\overline{\bm e}_{l:t}$ & Treatment history up to time t where intervention begins at period $l$ \\
$M_{ijt}(\overline{\bm z}_{t})$ & Potential intermediate variable under history $\overline{z}_{t}$. \\
$\text{PCE}_t(\overline{\bm z}_t, \overline{\bm z}_t^*; \mathcal{I})$ &  PCE comparing $\overline{\bm z}_t$ vs $\overline{\bm z}_t^*$ for stratum defined by region $\mathcal{I}$. \\
stPCE & Short-term PCE; at period $t$, it compares $\overline{\bm e}_{t:t}$ with $\overline{\bm 0}_{t}$. \\
ltPCE & Long-term PCE; at period $t$, it compares $\overline{\bm e}_{l:t}$ with $\overline{\bm 0}_{t}$, for any $l<t$. \\
\hline
\end{tabular}

\label{tab:notation}
\end{table}

{In the context of closed-cohort SW-CRTs, the \textit{short-term principal causal effect} (stPCE) in period $t$ is defined as the generalized PCE estimand that compares $\overline{\bm z}_t = \overline{\bm e}_{t:t}$ with $\overline{\bm z}_t^* = \overline{\bm 0}_t$. This estimand examines the immediate impact of treatment initiation during period $t$. Alternatively, the \textit{long-term principal causal effect} (ltPCE) assesses the cumulative impact after sustained exposure under the intervention. That is, at time $t$, the ltPCE under $t-l+1$ periods of sustained exposure involves comparing $\overline{\bm z}_t = \overline{\bm e}_{l:t}$ (with $l < t$) against $\overline{\bm z}_t^* = \overline{\bm 0}_t$. By differentiating the long-term effect estimand from the short-term effect estimand, our framework is sufficiently general to address potential effect heterogeneity due to differential exposure times, and generalizes the estimands definition in prior work \citet{KENNY2022,MALEYEFF2023,WANG2024} to accommodate an intermediate variable. The choice of estimands in practice depends on the scientific question of interest (studying the effect of initiation versus the effect sustained exposure), and the proposed identification strategy below allows the estimation of both stPCE and ltPCE. For ease of reference, we summarize the key notation in defining estimands in Table \ref{tab:notation}.}

\subsection{Identification under a sensitivity analysis framework}
\label{sec:assumption}

To point identify the PCEs,
we consider two additional structural assumptions that include interpretable sensitivity parameters. {Given the cluster-randomized nature of the SW-CRT, where randomization occurs at the cluster level potentially conditional on cluster-level covariates $\bm C_j$ (Assumption~\ref{assumption:randomization}), our structural assumptions are correspondingly defined conditional on $\bm C_j$. This implies that the sensitivity parameters characterizing the unobserved joint distributions depend on $\bm C_j$ but not on individual-level covariates $\bm X_{ij}$. Individual-level covariates $\bm X_{ij}$, however, can be accounted for in the working regression models (Section~\ref{sec:obs_data}) and are subsequently marginalized out when computing the PCEs.}

By construction of the principal causal estimands, we first need to identify the joint distribution of potential intermediate variables. Because the observed data do not provide any information on the joint, only their margins, we consider the following copula assumption to describe the joint \citep{EFRON1991,JIN2008,BARTOLUCCI2011,DANIELS2012}.

\begin{Assumption}[Copula for intermediate variables]
    \label{assumption:copula}
    For any time $t$ and any pair of treatment histories $\overline{\bm z}_t, \overline{\bm z}_t^*$, for any value of cluster-level covariates $\bm c$, we assume the joint distribution of $(M_{ijt}(\overline{\bm z}_t)$ and $M_{ijt}(\overline{\bm z}_t^*))$
    follows a copula family $\mathcal{\bm C}$ with sensitivity parameter, $\rho_{\bm c}$:
    \begin{align}\label{eq:copula}
        P(M_{ijt}(\overline{\bm z}_t), M_{ijt}(\overline{\bm z}_t^*)) 
        = \mathcal{C}_{\rho_{\bm c}}\left\{P(M_{ijt}(\overline{\bm z}_t)), P(M_{ijt}(\overline{\bm z}_t^*))\right\}.
    \end{align}
\end{Assumption}
For a continuous intermediate variable, a Gaussian copula is a common and practical choice due to its tractability and highly interpretable dependence parameter, which is a (rank) correlation parameter. The latter allows us to directly leverage the observed longitudinal correlations in a stepped wedge design for empirical calibration (Section~\ref{sec:parameter}). While other copulas (e.g., Clayton, Frank) offer different dependence structures, their parameters are generally less intuitive and harder to inform by the observed data. Therefore, we pursue the Gaussian copula approach in Assumption \ref{assumption:copula}. In its most general form, the correlation parameter could depend on the specific treatment histories being compared, i.e., $\rho_{\bm c}(\overline{\bm z}_t, \overline{\bm z}_t^*)$. For practical identifiability, we make a simplifying ``homogeneous'' assumption that the dependence structure, parameterized by  $\rho_{\bm c}$, is constant across all pairs of treatment regimens given cluster-level covariates. Assumption \ref{assumption:copula} allows us to construct the unobserved joint distribution from the observed marginals. In Section \ref{sec:parameter}, we will provide an approach to calibrate $\rho_{\bm c}$ using the observed data from a SW-CRT.

Identification of the PCE also requires specifying the distribution of the potential outcome within the principal strata. Inspired by \cite{HEAGERTY1999}, we propose the following identifying assumption. 
\begin{Assumption}[Marginal Structural Model]
    \label{assumption:msm}
    For any time $t$, pair of treatment histories $\overline{\bm z}_t, \overline{\bm z}_t^*$, cluster-level covariates value $c$, and sensitivity function $\lambda_{\bm c}(\cdot)$, the following holds:
    \begin{align}
    g\left\{E(Y_{ijt}(\overline{\bm z}_t)|M_{ijt}(\overline{\bm z}_t^*)=m^*, M_{ijt}(\overline{\bm z}_t)=m)\right\} = \Delta_{ijt}(m, \overline{\bm z}_t) + \lambda_{\bm c}(\overline{\bm z}_t - \overline{\bm z}_t^*) m^*.
    \label{eq:msm}
    \end{align}
\end{Assumption}

{
This assumption characterizes the relationship between a potential outcome under history $\overline{\bm z}_t$ and the pair of potential intermediate variables. This model is specifically marginal with respect to any latent cluster- or individual-level effects. The function $\Delta_{ijt}(m, \overline{\bm z}_t)$ is implicitly defined and can be recovered from the observed data through the marginal mean $E\{Y_{ijt}(\overline{\bm z}_t) \mid M_{ijt}(\overline{\bm z}_t)=m\}$ for a given $\lambda_{\bm c}(\cdot)$. In particular, when $\lambda_{\bm c}(\cdot)$ is nonzero, $\Delta_{ijt}(m, \overline{\bm z}_t)$ is obtained by solving a convolution equation (see the proof of Theorem \ref{thm:identification} and Section \ref{sec:identification}). The unidentified function $\lambda_{\bm c}(\overline{\bm z}_t - \overline{\bm z}_t^*)$ quantifies how the outcome $Y_{ijt}(\overline{\bm z}_t)$ is shifted (on the link-function scale) by the counterfactual intermediate variable $M_{ijt}(\overline{\bm z}_t^*)$, as a function of the contrast between the treatment histories. While in theory this shift could take many functional forms (e.g., additive, multiplicative, or based on other scales), its structure is not identified from the observed data. For tractability and ease of implementation, we adopt a simple linear form in our implementation. Furthermore, unlike approaches that rely on principal ignorability \citep{DING2017,jiang2022multiply,tong2025semiparametric}, our sensitivity analysis framework calibrates the untestable assumptions using features of the stepped wedge design (see Section \ref{sec:parameter}). This makes the proposed approach particularly relevant in settings such as our motivating trial, where only a limited set of baseline covariates is available.
With our structural assumptions, we can achieve point identification as follows.
}

\begin{Theorem}[Identification]
    \label{thm:identification}
    Under Assumptions \ref{assumption:sutva}–\ref{assumption:msm}, the estimands in \eqref{eq:estimand} are point-identified for any pair of treatment histories $\overline{\bm z}_t, \overline{\bm z}_t^*$:
        \begin{align*}
        &\text{PCE}_t(\overline{\bm z}_t, \overline{\bm z}_t^*; \mathcal{I}) = \\
        &\frac{
        \int_{\mathcal{I}} \left[ g^{-1}(\Delta_{ijt}(m, \overline{\bm z}_t) + \lambda_c(\overline{\bm z}_t - \overline{\bm z}_t^*) m^*) - g^{-1}(\Delta_{ijt}(m^*, \overline{\bm z}_t^*) + \lambda_{\bm c}(\overline{\bm z}_t^* - \overline{\bm z}_t) m) \right] dP(m, m^*)
        }
        {
        \int_{\mathcal{I}} dP(m, m^*)
        },
    \end{align*}
    where the integration is over the joint distribution $P(M_{ijt}(\overline{\bm z}_t)=m, M_{ijt}(\overline{\bm z}_t^*)=m^*)$.
\end{Theorem}

Supplementary Section B provides a proof for this identification result. The essential ingredients of the identification formula are the marginal structural model \eqref{eq:msm} and the joint distribution of the intermediate variables derived from the copula structure \eqref{eq:copula}. Identification is conditional on sensitivity parameter $\rho_{\bm c}$ and sensitivity function $\lambda_{\bm c}(\cdot)$.

\section{Model specification, Sensitivity parameters, and Bayesian Inference}\label{sec:model}

\subsection{Observed data model}
\label{sec:obs_data}

To estimate the $\text{PCE}_t(\overline{\bm z}_t, \overline{\bm z}_t^*; \mathcal{I})$, we need an observed data model that allows the treatment effects to vary with duration of exposure. This aligns with \citet{WANG2024, KENNY2022, MALEYEFF2023}, who emphasize the critical importance in flexible modeling of the treatment effect curve to ensure robustness. We build upon the widely used mixed-effects model framework \citep{LI2021} and provide an extension to explicitly parameterize the effect of treatment duration in our context with an intermediate variable. Let $D_{jt} = \sum_{k=1}^t Z_{jk}$ be the cumulative number of periods that cluster $j$ has been exposed to the intervention at time $t$. For the continuous intermediate variable and binary outcome in our application, we consider the following models based on the duration variable:
\begin{align}
    M_{ijt} &= \eta_{1t} + \sum_{d=1}^{t} \gamma_d I(D_{jt}=d) + {\bm X}_{ij}^\top{\bm \omega}_1 + {\bm C}_{j}^\top{\bm \omega}_2 + \alpha_{1j} + \phi_{1ij} + \epsilon_{ijt}, \label{eq:reg M} \\
    \logit(\mu_{Y,ijt}) &= \eta_{2t} + \sum_{d=1}^{t} \beta_d I(D_{jt}=d) + {\bm X}_{ij}^\top{\bm \psi}_1 + {\bm C}_{j}^\top{\bm \psi}_2 + M_{ijt}{\psi}_3 + M_{ijt}\psi_4 I(D_{jt}>0) + \alpha_{2j} + \phi_{2ij}. \label{eq:reg Y}
\end{align}
Here, $I(D_{jt}=d)$ is an indicator function for the treatment duration being exactly $d$ periods. The parameters $\gamma_d$ and $\beta_d$ represent the marginal effects of a $d$-period-long treatment exposure at time $t$, relative to the control condition ($D_{jt}=0$). The models also adjust for baseline cluster- and individual-level covariates. The random effects assumption includes cluster-level intercepts, $(\alpha_{1j},\alpha_{2j}) \sim N(\bm{0}, \bm\Sigma_{\bm\alpha})$ and individual-level intercepts $(\phi_{1ij},\phi_{2ij}) \sim N(\bm{0}, \bm\Sigma_{\bm \phi})$, accounting for between-individual and within-individual dependency structures. The covariance matrices $\bm \Sigma_{\bm \alpha}$ and $\bm \Sigma_{\bm \phi}$ represent the additional association between the intermediate variable and the outcome at both the cluster and individual levels.

\subsection{Identification}
\label{sec:identification}

To connect these observed data models to the identification result in Theorem \ref{thm:identification}, we first link the model parameters to the potential outcome distributions. Under model \eqref{eq:reg M}, the potential intermediate outcome $M_{ijt}(\overline{\bm z}_t)$ for any history $\overline{\bm z}_t$ follows a normal distribution. Let $d_z = \sum_{k=1}^t z_k$ be the duration of treatment under history $\overline{\bm z}_t$. The mean and variance are $E(M_{ijt}(\overline{\bm z}_t)) = \eta_{1t} + \gamma_{d_z} + \bm{X}_{ij}^\top{\bm \omega}_1 + \bm{C}_{j}^\top{\bm\omega}_2$ (where $\gamma_0=0$) and $\text{Var}(M_{ijt}(\overline{\bm z}_t)) = \Sigma_{\alpha,11} + \Sigma_{\phi,11} + \sigma_\epsilon^2$, where $\Sigma_{\alpha,ll'}$ and $\Sigma_{\phi,ll'}$ are the $(l,l')$th entries of $\bm\Sigma_{\bm\alpha}$ and $\bm\Sigma_{\bm \phi}$. Therefore, for a given sensitivity parameter $\rho_{\bm c}$, Assumption \ref{assumption:copula} allows us to calculate the denominator in Theorem \ref{thm:identification}. That is, for any two histories $\overline{\bm z}_t$ and $\overline{\bm z}_t^*$, the difference $M_{ijt}(\overline{\bm z}_t) - M_{ijt}(\overline{\bm z}_t^*)$ is normally distributed with mean $\gamma_{d_z} - \gamma_{d_{z^*}}$ and variance $2(1-\rho)(\Sigma_{\alpha,11} + \Sigma_{\phi,11} + \sigma_\epsilon^2)$. If the subpopulation is defined by $\mathcal{I}=[a,b]$, the denominator probability is:
\begin{align*}
    \int_{\mathcal{I}}
        &dP(M_{ijt}(\overline{\bm z}_t),M_{ijt}(\overline{\bm z}_t^*))
        = P(a \leq M_{ijt}(\overline{\bm z}_t) - M_{ijt}(\overline{\bm z}_t^*) < b) \\
        =& \Phi\left(\frac{b-(\gamma_{d_z} - \gamma_{d_{z^*}})}{\sqrt{2(1-\rho)(\Sigma_{\alpha,11} + \Sigma_{\phi,11} + \sigma_\epsilon^2)}}\right) - \Phi\left(\frac{a-(\gamma_{d_z} - \gamma_{d_{z^*}})}{\sqrt{2(1-\rho)(\Sigma_{\alpha,11} + \Sigma_{\phi,11} + \sigma_\epsilon^2)}}\right).
\end{align*}

To compute the numerator in Theorem \ref{thm:identification}, we first need an expression for the conditional expectation $E(Y_{ijt}(\overline{\bm z}_t)|M_{ijt}(\overline{\bm z}_t)=m)$ based on our observed data models. Given the joint normality of the random effects and the error term for $M_{ijt}$, the conditional distribution $(\alpha_{2j} + \phi_{2ij})|M_{ijt}(\overline{\bm z}_t)=m$ is also normal, with mean and variance
\begin{align*}
    E(\alpha_{2j} + \phi_{2ij}|M_{ijt}(\overline{\bm z}_t)=m) 
        &= \frac{\Sigma_{\alpha,12} + \Sigma_{\phi,12}}{\Sigma_{\alpha,11} + \Sigma_{\phi,11} + \sigma^2_{\epsilon}}(m - E(M_{ijt}(\overline{\bm z}_t))), \\
    \text{Var}(\alpha_{2j} + \phi_{2ij}|M_{ijt}(\overline{\bm z}_t)=m)
        &= \Sigma_{\alpha,22} + \Sigma_{\phi,22} - \frac{(\Sigma_{\alpha,12} + \Sigma_{\phi,12})^2}{\Sigma_{\alpha,11} + \Sigma_{\phi,11} + \sigma^2_{\epsilon}},
\end{align*}
where $E(M_{ijt}(\overline{\bm z}_t))$ is the marginal mean. 
We can then compute $E(Y_{ijt}|M_{ijt}=m, \overline{\bm Z}_{jt}=\overline{\bm z}_t) = E\left\{\logit^{-1}(\eta_{2t} + \beta_{d_z} + \dots + \alpha_{2j} + \phi_{2ij})\right\}$ by integrating out the random effects using Gaussian-Hermite quadrature.

The function $\Delta_{ijt}(m, \overline{\bm z}_t)$ from Assumption \ref{assumption:msm} is identified by solving the following convolution equation (see Supplementary Section B) for each pair of treatment histories: 
\begin{align}
    &E(Y_{ijt}|M_{ijt} = m, \overline{\bm Z}_{jt} = \overline{\bm z}_t) \\  
    =& \int g^{-1}(\Delta_{ijt}(m,\overline{\bm z}_t) + \lambda_{\bm c}(\overline{\bm z}_t - \overline{\bm z}_t^*) m^*) dP(M_{ijt}(\overline{\bm z}_t^*)|M_{ijt}(\overline{\bm z}_t) = m). \nonumber
\end{align}
We use the Newton-Raphson algorithm to solve for $\Delta_{ijt}(m, \overline{\bm z}_t)$; the details are provided in the supplement. Once we solve for $\Delta_{ijt}(m, \overline{\bm z}_t)$ for any $m$ and history $\overline{\bm z}_t$, using Assumption \ref{assumption:msm}, the conditional expectation of the potential outcome for any pair of potential intermediate values is
\begin{align*}
    E\left(Y_{ijt}(\overline{\bm z}_t)|M_{ijt}(\overline{\bm z}_t\right)=m, M_{ijt}(\overline{\bm z}_t^*)=m^*) = \logit^{-1}\left(\Delta_{ijt}(m, \overline{\bm z}_t) + \lambda_{\bm c}(\overline{\bm z}_t - \overline{\bm z}_t^*) m^*\right).
\end{align*}
With this expectation expression, we can compute the numerator of the PCE in Theorem \ref{thm:identification} using Monte Carlo integration; additional details are provided in the supplement.

\subsection{Calibration of sensitivity parameters}
\label{sec:parameter}

The identification formula in Theorem \ref{thm:identification} requires specification of  sensitivity parameters. While the theoretical framework developed in Section \ref{sec:methods} is general enough to accommodate any pair of treatment histories, we leverage the unique structure of SW-CRTs to calibrate sensitivity parameters for estimands comparing any treatment history $\overline{\bm z}_t$ against the full control history $\overline{\bm 0}_t$. These estimands include both the short-term (stPCE) and long-term (ltPCE) principal causal effects. As established in Section \ref{sec:methods}, the identification relies on Assumptions \ref{assumption:copula} and \ref{assumption:msm} which are conditional on cluster-level covariates $\bm C_j$, but marginalized over individual-level covariates $\bm X_{ij}$. 

For Assumption \ref{assumption:copula}, $\rho_{\bm c}$ characterizes the correlation between potential intermediate outcomes, conditional on $\bm C_j$. For a contrast between $\overline{\bm z}_t$ and $\overline{\bm 0}_t$, $\rho_{\bm c}$ describes the correlation between $M_{ijt}(\overline{\bm z}_t)$ and $M_{ijt}(\overline{\bm 0}_t)$ given $\bm C_j=\bm c$. We leverage the staggered rollout and longitudinal data of SW-CRT designs to inform this parameter. We use the observed correlation between adjacent time points under the intervention (e.g., $M_{ij,t-1}$ and $M_{ijt}$) as a conservative estimate (lower bound) for the unobserved cross-world correlation $\rho_{\bm c}$. We assume this correlation is duration-independent and does not depend on $d$.

To estimate this correlation $\rho_{\bm c}^*$ marginalized over $\bm X_{ij}$, we must account for the dependence induced by $\bm X_{ij}$ on both mediators. We utilize data pooled from all periods $t$ where the intervention is active ($Z_{jt}=1$) and fit auxiliary linear regression models. To isolate the residual correlation, these models must adjust for the effect of duration $D_{jt}$ on the mean structure, along with $\bm X_{ij}$ and $\bm C_j$:
\begin{equation}
\begin{aligned}
M_{ij,t-1} &= \alpha_0(D_{jt}) + \bm X_{ij}^\top\bm \alpha_{\bm X} + \bm C_j^\top\bm \alpha_{\bm C} + \epsilon_{ij,t-1} \\
M_{ijt} &= \beta_0(D_{jt}) + \bm X_{ij}^\top\bm\beta_{\bm X} + \bm C_j^\top\bm\beta_{\bm C} + \epsilon_{ijt}.
\end{aligned}
\label{eq:auxiliary_rho_models}
\end{equation}
We then apply the Law of Total Covariance to obtain the moments marginalized over $\bm X_{ij}$ (conditional on $\bm C_j$), assuming the residual covariance structure is constant across durations. The marginalized covariance is:
\begin{align*}
\text{Cov}(M_{ij,t-1}, M_{ijt} | \bm C_j) &= E_{\bm X}\left\{\text{Cov}(\cdot | \bm X_{ij}, \bm C_j) |\bm C_j\right] + \text{Cov}_{\bm X}\left[E(\cdot|\bm X_{ij}, \bm C_j) | C_j\right] \\
&= \text{Cov}(\epsilon_{ij,t-1}, \epsilon_{ijt}) + \bm\alpha_{\bm X}^\top \text{Var}(\bm X_{ij}|\bm C_j) \bm\beta_{\bm X}.
\end{align*}
Similarly, the marginalized variances are:
\begin{align*}
\text{Var}(M_{ij,t-1} | \bm C_j) &= \text{Var}(\epsilon_{ij,t-1}) + \bm \alpha_{\bm X}^\top \text{Var}(\bm X_{ij}|\bm C_j) \bm \alpha_{\bm X} \\
\text{Var}(M_{ijt} | \bm C_j) &= \text{Var}(\epsilon_{ijt}) + \bm \beta_{\bm X}^\top \text{Var}(\bm X_{ij}|\bm C_j) \bm \beta_{\bm X}.
\end{align*}
This requires estimating the regression parameters and the residual covariances from \eqref{eq:auxiliary_rho_models}, and the covariance matrix of $\bm X_{ij}$ conditional on $\bm C_j$. {Furthermore, since $\bm C_j$ is categorical in our setting, $\text{Var}(\bm X_{ij}\mid \bm C_j)$ is estimated by the pooled empirical covariance of $\bm X_{ij}$ within strata defined by $\bm C_j$. If $\bm C_j$ were continuous, one could analogously estimate $\text{Var}(\bm X_{ij}\mid \bm C_j)$ via mild binning of $\bm C_j$ or other local smoothing.}
The marginalized correlation $\rho_{\bm c}^*$ is then estimated as:
\begin{align}
    \label{eq:rho}
    \rho_{\bm c}^*
    &=
    \frac{
        \widehat{\text{Cov}}(M_{ij,t-1}, M_{ijt} | \bm C_j)
    }
    {
        \sqrt{\widehat{\text{Var}}( M_{ij,t-1} | \bm C_j)}
        \sqrt{\widehat{\text{Var}}( M_{ijt} | \bm C_j)}
    }.
\end{align}
We expect that this observed longitudinal correlation to serve as a conservative estimate for the unobserved cross-world correlation $\rho_{\bm c}$. 

For Assumption \ref{assumption:msm}, we need to specify the sensitivity function $\lambda_{\bm c}(\cdot)$. Specifically, we focus on $\lambda_{\bm c}(d) \equiv \lambda(\overline{\bm z}_t - \overline{\bm 0}_t|\bm C_j)$ and $\lambda_{\bm c}(-d) \equiv \lambda(\overline{\bm 0}_t - \overline{\bm z}_t|\bm C_j)$, where $d$ is the duration associated with history $\overline{\bm z}_t$. We leverage the longitudinal structure of the SW-CRT to obtain empirically informed distributions for these unobserved, cross-world associations.

Similar to the sensitivity parameter calibration in Assumption \ref{assumption:copula}, we leverage the longitudinal observations to define auxiliary parameters based on observable proxies. We allow these associations to vary by exposure duration $d$. Specifically, we fit flexible auxiliary regression models using data from all treated periods, modeling the interaction between mediators and duration $D_{jt}=d$, and include individual and cluster-level covariates:
\begin{equation}
\begin{aligned}
g\left\{E(Y_{ijt} | D_{jt}=d, M_{ijt}, M_{ij,t-1}, \bm X_{ij}, \bm C_j)\right\} &= \theta_0(d) + \theta_1(d) M_{ijt} + \theta_2(d) M_{ij,t-1} \\
&+ \bm X_{ij}^\top \bm \theta_{\bm X} + \bm C_j^\top \bm \theta_{\bm C} \\
g\left\{E(Y_{ij,t-1} | D_{jt}=d, M_{ijt}, M_{ij,t-1}, \bm X_{ij}, \bm C_j)\right\} &= \zeta_0(d) + \zeta_1(d) M_{ijt} + \zeta_2(d) M_{ij,t-1} \\
&+ \bm X_{ij}^\top \bm \zeta_{\bm X} + \bm C_j^\top \bm \zeta_{\bm C}.
\end{aligned}
\label{eq:auxiliary_models_general}
\end{equation}
We aim to obtain the association marginalized over $\bm X_{ij}$. If $g(\cdot)$ is the identity link (linear model) and the specification is additive (no interactions between mediators and $\bm X_{ij}$), the conditional associations $\theta_2(d)$ and $\zeta_1(d)$ are equal to the marginalized associations. If $g(\cdot)$ is a non-linear link, explicit marginalization is generally required. However, in the context of sensitivity analysis calibration, we use the conditional coefficients as approximations for the marginalized parameters.

We treat the posterior distributions of $\theta_2(d)$ and $\zeta_1(d)$ as the calibrated distributions for $\lambda_{\bm c}(d)$ and $\lambda_{\bm c}(-d)$, respectively. This calibration relies on the assumption that the observed lagged associations (e.g., the effect of $M_{ij,t-1}$ on $Y_{ijt}$) approximate the unobserved cross-world associations (e.g., the effect of $M_{ijt}(\overline{\bm 0}_t)$ on $Y_{ijt}(\overline{\bm z}_t)$).

However, it is possible that the calibrated values derived from lagged effects underestimate the true strength of the contemporaneous counterfactual associations. To address this uncertainty, we define the sensitivity parameters as: $\lambda_{\bm c}(d) = k \cdot \theta_2(d)$, $\lambda_{\bm c}(-d) = k \cdot \zeta_1(d)$, where $k$ is a scaling factor. The baseline analysis corresponds to $k=1.0$, utilizing the calibrated distributions directly. We then conduct sensitivity analyses by varying $k$ (e.g., $k=0.5, 1.5, 2.0$) to examine how the PCE estimates change if the true association strength is different from the calibrated values.

To incorporate the uncertainty associated with the calibration, we adopt a fully Bayesian approach. We fit the auxiliary models for $\rho_{\bm c}^*$ (using models \eqref{eq:auxiliary_rho_models} and \eqref{eq:rho}) and the calibrated $\lambda$ values ($\theta_2(d), \zeta_1(d)$ from models \eqref{eq:auxiliary_models_general}) using Bayesian methods via \texttt{RStan}. This procedure yields posterior distributions for these calibrated parameters, allowing their uncertainty to propagate into the PCE estimation.

\subsection{Bayesian inference and posterior computation}\label{sec:posterior}

We specify weakly informative priors for each model parameter \eqref{eq:reg M} and \eqref{eq:reg Y}. Details can be found in the supplement. To generate posterior samples, we use Hamiltonian Monte Carlo (HMC; \cite{NEAL2011}) via the \texttt{Stan} software package \citep{Carpenter2017}.

When calculating the posterior of PCEs, we integrate the sensitivity parameter calibration stochastically into the Bayesian workflow. For the duration-dependent sensitivity functions $\lambda_{\bm c}(d)$ and $\lambda_{\bm c}(-d)$, we utilize the calibrated distributions derived in Section \ref{sec:parameter}. In each iteration, we draw realizations of the calibrated parameters ($\theta_2(d), \zeta_1(d)$) from their respective posterior distributions obtained from the auxiliary Bayesian models. We then apply a set of scaling factors. The baseline analysis uses $k=1.0$. For each $k$, we calculate the sensitivity parameters as $\lambda_{\bm c}(d) = k \cdot \theta_2(d)$ and $\lambda_{\bm c}(-d) = k \cdot \zeta_1(d)$ using the sampled realizations. 

For the duration-independent correlation $\rho_{\bm c}$, we utilize the posterior distribution of the calibrated value $\rho_{\bm c}^*$ obtained from the auxiliary Bayesian model (estimated via \eqref{eq:rho}). In the `Calibrated' scenario (used in the primary analysis), we sample $\rho_{\bm c}$ directly from this posterior distribution in each iteration. We also conduct sensitivity analysis by fixing $\rho_{\bm c}$ at specific values (e.g., 0.8 or 0.9).

Algorithm \ref{alg:PCE} presents the complete procedure for obtaining posterior samples of the PCE. The process involves fitting the observed data model (Step 1), fitting the Bayesian auxiliary models to obtain posterior distributions for the calibrated parameters (Step 2), and computing the PCEs while integrating over the uncertainty of both the model parameters and the sensitivity parameters across different scaling factors $k$ and $\rho_{\bm c}$ scenarios (Step 3). These posterior samples of PCEs enable the calculation of credible intervals and other posterior summaries.

\begin{algorithm}[ht]
\caption{PCE Estimation with Sensitivity Analysis}
\label{alg:PCE}
\KwIn{
Estimand: $\text{PCE}_t(\overline{\bm e}_{l:t}, \overline{\bm 0}_t; \mathcal{I})$; Duration $d=t-l+1$;
Observed data; Scaling factors $K$ for $\lambda$; Scenarios for $\rho$;
Posterior sample size $S$.
}

\SetKwBlock{StepOne}{Step 1. Observed Data Model Fitting (MCMC)}{end}
\SetKwBlock{StepTwo}{Step 2. Calibration (Bayesian Auxiliary Models)}{end}
\SetKwBlock{StepThree}{Step 3. PCE Computation and Sensitivity Analysis}{end}

\StepOne{
Specify priors for parameters $\bm{\xi}$ of models \eqref{eq:reg M} and \eqref{eq:reg Y}. Use \texttt{Stan} to obtain posterior samples $\{\bm{\xi}^{(s)}\}_{s=1}^S$.
}

\StepTwo{
Fit Bayesian auxiliary models to obtain posterior distributions for the calibrated parameters: $\rho_{\bm c}^*$ (using \eqref{eq:auxiliary_rho_models} and \eqref{eq:rho}), and $\theta_2(d), \zeta_1(d)$ (using \eqref{eq:auxiliary_models_general}).
}

\StepThree{
\For{each Sensitivity Parameter ($\lambda$, $\rho$)}{
    \For{$s \gets 1$ \KwTo $S$} {
        1. Select posterior sample $\bm{\xi}^{(s)}$\;
        2. Sample Calibrated Parameters: Draw one realization from the posterior distributions for $\rho_{\bm c}^{*(s)}$, $\theta_2^{(s)}(d)$, and $\zeta_1^{(s)}(d)$\;
        3. Compute Sensitivity Parameters:\;
        \hspace{1em} $\bullet$ $\rho^{(s)}$: Set according to the current $\rho$ scenario (e.g., use $\rho_{\bm c}^{*(s)}$ or fixed value).\;
        \hspace{1em} $\bullet$ $\lambda^{(s)}(d) = k \cdot \theta_2^{(s)}(d)$; $\lambda^{(s)}(-d) = k \cdot \zeta_1^{(s)}(d)$.\;
        4. Compute PCE: Solve for $\Delta_{ijt}(\cdot)$ and calculate $\text{PCE}_t^{(s)}$ using Theorem \ref{thm:identification} with $\bm{\xi}^{(s)}, \rho^{(s)}, \lambda^{(s)}(\cdot)$\;
        5. Store $\text{PCE}_t^{(s)}$ for the current $k$ and $\rho$\;
    }
}
}
\KwOut{Posterior samples of PCEs for all settings of the sensitivity parameters.}
\end{algorithm}

\subsubsection{Missing Data}\label{sec:missing}
In our application, we have missing data in both outcomes and intermediate variables which is only due to dropout (and thus is exclusively monotone). Define $\bm U_{ijt} = (M_{ijt}, Y_{ijt})$ and define an observed data indicator $R_{ijt} = I\{\bm U_{ijt} \mbox{ is observed}\}$. We assume 
{
$$R_{ijt} \perp  \bm U_{ijt},\ldots,\bm U_{ijT} | \bm U_{ij1},\ldots,\bm U_{ijt-1}, \overline{\bm Z}_{jt}, \bm X_{ij}, \bm C_j,$$
}
\noindent i.e., missing at random dropout.  We discuss weakening this assumption in Section 6.

\section{Analysis of the Crowdsourcing HIV Intervention Study}\label{sec:analysis}

This section presents the application of our proposed methodology to the HIV crowdsourcing study. The study design included only participants who had not tested for HIV in the past three months. Consequently, our analysis focuses on time points after the first observation ($t > 1$). Regarding dropout, we note that both the outcome ($Y$) and the intermediate variable ($M$) are always missing simultaneously (monotone missingness). Dropouts are addressed under an assumption of ignorable missingness as described in Section~\ref{sec:missing}.

We use \texttt{RStan} to sample from the posterior distribution of the observed data model parameters. We then compute the following causal estimands by utilizing these posterior samples along with the identifying assumptions. We focus on both short-term (stPCE) and long-term (ltPCE) principal causal effects, as introduced in Section~\ref{sec:PO}. The stPCE at time $t$ compares immediate intervention initiation $(\overline{\bm e}_{t:t})$ versus control $(\overline{\bm 0}_{t})$:
$\text{stPCE}_*(t)=E[Y_{ijt}(\overline{\bm e}_{t:t})-Y_{ijt}(\overline{\bm 0}_{t})|M_{ijt}(\overline{\bm e}_{t:t})-M_{ijt}(\overline{\bm 0}_{t})\in\mathcal{I}_{*}]$. 
To examines the impact of sustained exposure, we define the ltPCE at time $t$ for duration $d$ ($t>d$) by comparing sustained exposure $(\overline{\bm e}_{t-d+1:t})$ versus control $(\overline{\bm 0}_{t})$:
$\text{ltPCE}_{*}(t, d)=E[Y_{ijt}(\overline{\bm e}_{t-d+1:t})-Y_{ijt}(\overline{\bm 0}_{t})|M_{ijt}(\overline{\bm e}_{t-d+1:t})-M_{ijt}(\overline{\bm 0}_{t})\in\mathcal{I}_{*}]$. We report results for durations up to $d=3$.

\begin{figure}[ht!]
    \centering
    \includegraphics[width=\textwidth]{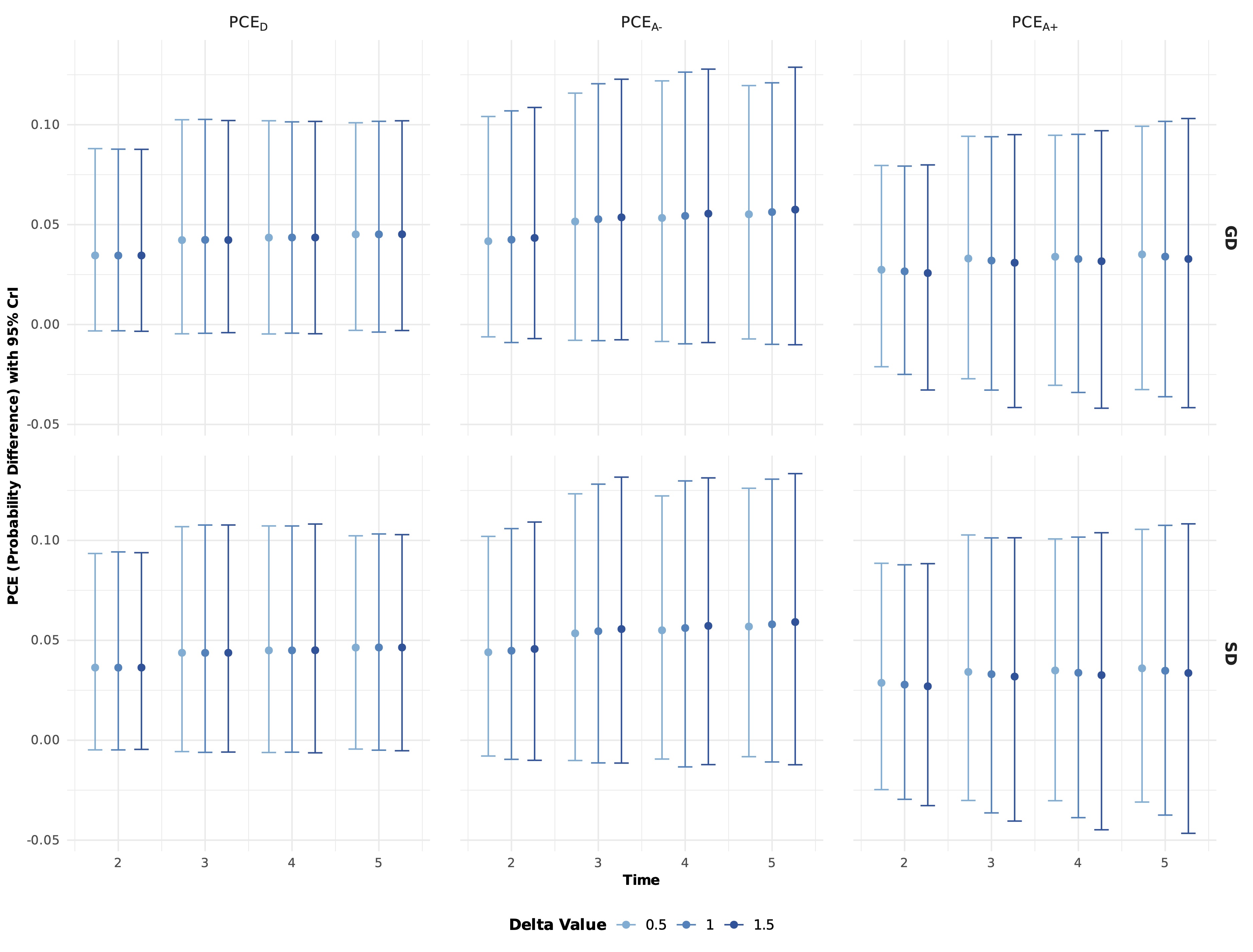}
    \caption{Sensitivity analysis of stPCE ($d=1$) to the indifference threshold $\delta$. Results are sorted by Stratum (Columns) and Province (Rows). Colored lines represent different $\delta$ values (0.5, 1.0, 1.5). The analysis uses the primary settings for $\rho_{\bm c}$ (calibrated) and $k=1.0$ for $\lambda_{\bm c}(\cdot)$.}    \label{fig:PCE_delta_sensitivity}
\end{figure}

Recall that $Y_{ijt}$ is HIV testing uptake and $M_{ijt}$ is perceived social norms. For the primary analysis, we define three intervals based on a threshold $\delta=0.5$: $\mathcal{I}_{\text{D}} = [-0.5, 0.5]$ (dissociative stratum), $\mathcal{I}_{\text{A}-} = (-\infty, -0.5)$ (negative associative  stratum), and $\mathcal{I}_{\text{A}+} = (0.5, \infty)$ (positive associative  stratum). We also conduct sensitivity analyses using $\delta=1.0$ and $\delta=1.5$. The intermediate variable ranges from 6 to 24, derived from the sum of six items scored on a 1-4 scale. The primary threshold of $\delta=0.5$ represents a relatively small magnitude of change (e.g., less than a one-point shift on a single item on average). We selected this threshold to define the dissociative stratum rigorously, while sensitivity analyses (Figure \ref{fig:PCE_delta_sensitivity}) explore robustness to this choice. $\text{PCE}_{\text{D}}$ represents the dissociative effect, estimating the intervention's impact for participants with minimal change in social norms. In contrast, $\text{PCE}_{\text{A}-}$ and $\text{PCE}_{\text{A}+}$ capture associative effects, focusing on participants with a decrease ($< -0.5$) and increase ($> 0.5$) in social norms, respectively. These PCEs allow us to examine treatment effect heterogeneity across these distinct subgroups defined by the potential responses of the intermediate variable.

The sensitivity parameters, $\lambda_{\bm c}(\cdot)$ and $\rho_{\bm c}$, were handled using  calibration and scaling factors as described in Section~\ref{sec:parameter} and \ref{sec:posterior}. The analysis marginalizes over individual-level covariates ($\bm X_{ij}$) but is presented conditional on the cluster-level covariate, Province ($C_j$: Guangdong [GD] and Shandong [SD]). Our primary analysis utilizes the baseline scaling factor $k=1.0$ for $\lambda_{\bm c}(\cdot)$ and the 'Calibrated' scenario for $\rho_{\bm c}$ (using the posterior distribution of the duration-independent $\rho^{*}$). We conduct extensive sensitivity analyses using different scaling factors for $\lambda_{\bm c}(\cdot)$ ($k=0.5, 1.5, 2.0$) and fixed values for $\rho_{\bm c}$ (0.8 and 0.9).

\begin{figure}[ht!]
    \centering
    \includegraphics[width=\textwidth]{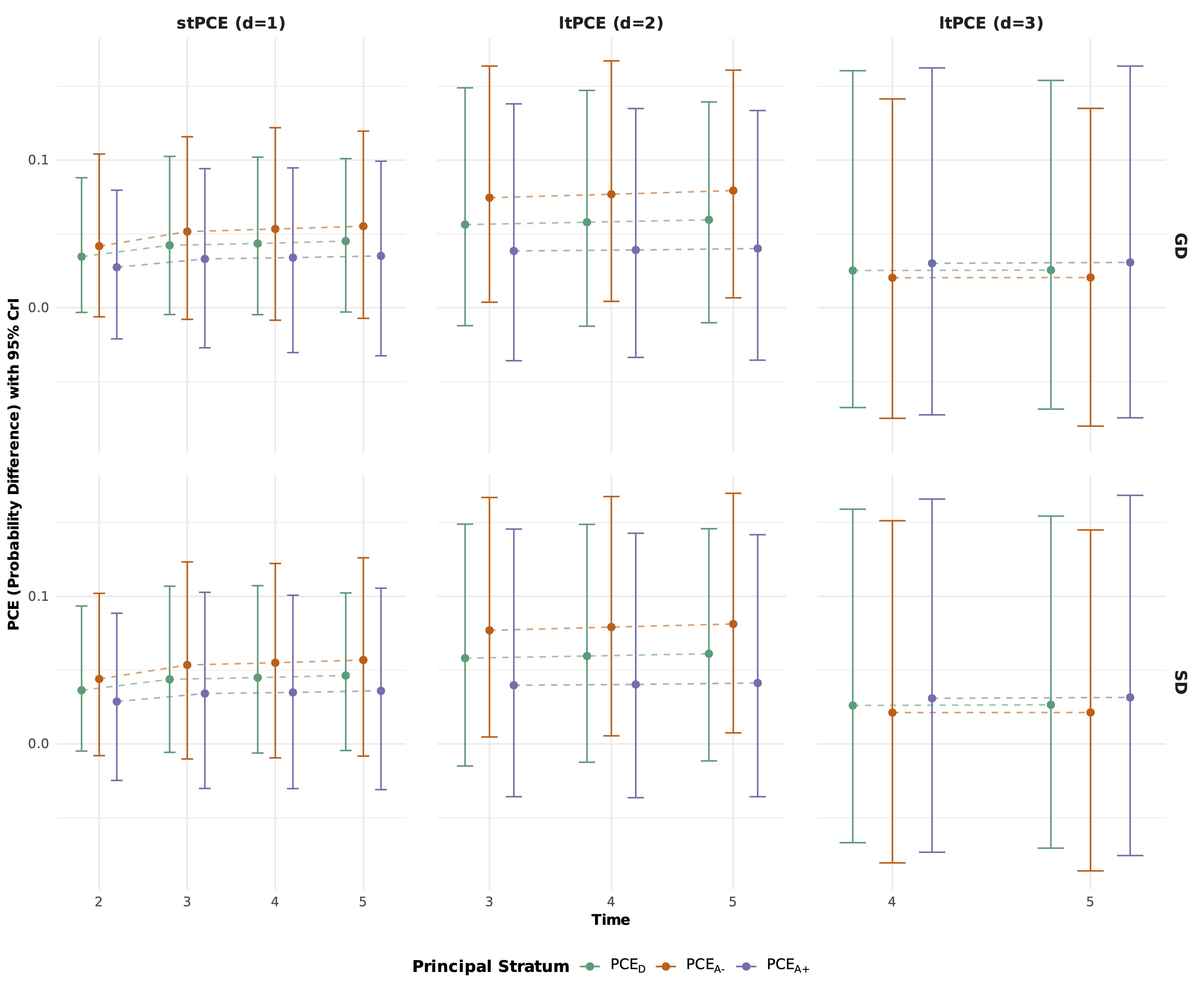}
    \caption{Posterior summaries (mean and 95\% credible intervals) of $\text{PCE}_{\text{D}}$ (Dissociative), $\text{PCE}_{\text{A}-}$ (Associative Negative), and $\text{PCE}_{\text{A}+}$ (Associative Positive) across time periods with $\delta=0.5$. Results are conditional on Province (Rows: GD, SD) and sorted by Duration (Columns: stPCE with $d=1$, ltPCE with $d=2$, ltPCE with $d=3$). This primary analysis uses the 'Calibrated' scenario for $\rho_{\bm c}$ and the baseline scaling factor $k=1.0$ for $\lambda_{\bm c}(\cdot)$.}    \label{fig:PCE_primary}
\end{figure}

Figure \ref{fig:PCE_primary} presents the posterior means and 95\% credible intervals of the PCEs for the primary analysis ($\delta=0.5, k=1.0$). The results indicate a potential positive effect of the intervention on HIV testing uptake across all principal strata, durations, and provinces, with posterior means generally ranging from 0.03 to 0.07. However, most of the 95\% credible intervals generally include zero. Furthermore, the magnitudes of the estimated effects appear similar across the three principal strata ($\mathcal{I}_{\text{D}}$, $\mathcal{I}_{\text{A}-}$, and $\mathcal{I}_{\text{A}+}$). While point estimates might suggest subtle differences, the 95\% credible intervals exhibit substantial overlap within every panel. Therefore, we did not observe strong evidence of treatment effect heterogeneity based on how the intervention potentially affected an individual's perceived social norms.

\begin{figure}[ht!]
    \centering
    \includegraphics[width=\textwidth]{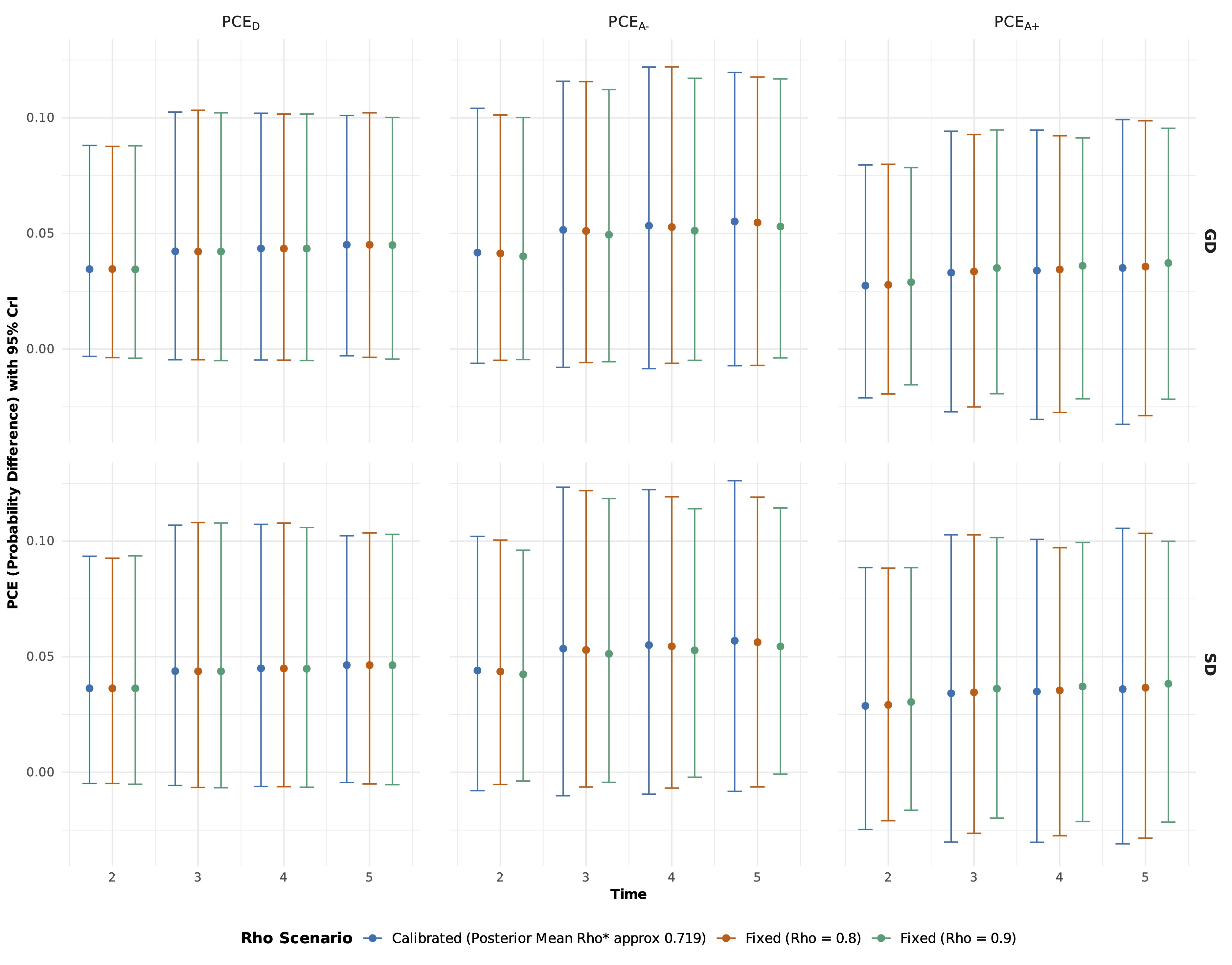}
    \caption{Sensitivity analysis of stPCE ($d=1$) to the correlation coefficient scenario for $\rho_{\bm c}$ with $\delta=0.5$ and $k=1.0$. Results are sorted by Stratum (Columns) and Province (Rows). Colored lines represent the different $\rho_{\bm c}$ scenarios.}    \label{fig:PCE_rho_sensitivity}
\end{figure}

We conducted comprehensive sensitivity analyses to assess the impact of the sensitivity parameters ($\rho_{\bm c}$ and $\lambda_{\bm c}(\cdot)$) and the indifference threshold $\delta$. Figure \ref{fig:PCE_rho_sensitivity} illustrates the sensitivity to $\rho_c$ for the stPCE with $d=1$, comparing the calibrated scenario (where the posterior mean of $\rho^*$ is 0.719) against fixed values of 0.8 and 0.9 (with $\delta=0.5$ and $k=1.0$). The results demonstrate robustness to this sensitivity parameter. The posterior means and credible intervals are nearly identical across the different scenarios.

\begin{figure}[ht!]
    \centering
    \includegraphics[width=\textwidth]{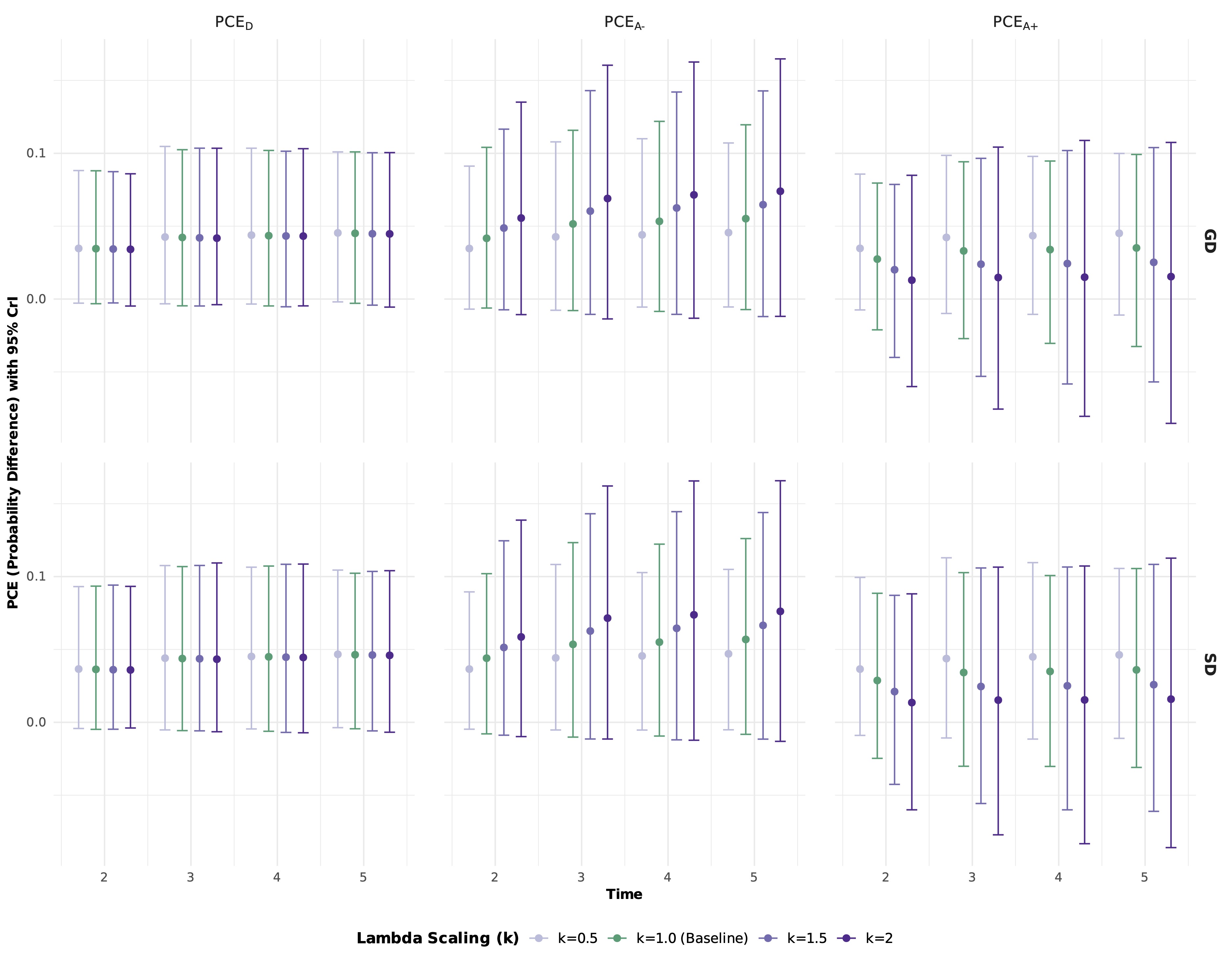}
    \caption{Scaling Factor Sensitivity Analysis of stPCE ($d=1$) to the scaling factor $k$ for $\lambda_{\bm c}(\cdot)$, with $\delta=0.5$ and calibrated $\rho_{\bm c}$. Results are sorted by Stratum (Columns) and Province (Rows). Colored lines represent the different scaling factors $k$.}
    \label{fig:PCE_lambda_sensitivity}
\end{figure}

Furthermore, we examine the sensitivity to the scaling factor $k$ for $\lambda_{\bm c}(\cdot)$ in Figure \ref{fig:PCE_lambda_sensitivity}, comparing the baseline ($k=1.0$) with scaled scenarios ($k=0.5, 1.5, 2.0$). The results show moderate sensitivity to this parameter, and interestingly, the direction of sensitivity varies by principal stratum. For the Negative Associative stratum ($\mathcal{I}_{\text{A}-}$), the estimated effects increase as $k$ increases. Conversely, for the Positive Associative stratum ($\mathcal{I}_{\text{A}+}$), the estimated effects decrease as $k$ increases. The Dissociative stratum ($PCE_1$) shows minimal sensitivity to $k$.

This differential sensitivity means that as $k$ increases—representing a stronger assumed association between the counterfactual mediator and the outcome (Assumption \ref{assumption:msm})—the estimates across the strata diverge. For example, when $k=2.0$, the posterior mean for $\text{PCE}_{\text{A}-}$ is noticeably higher, while the posterior mean for $\text{PCE}_{\text{A}+}$ is noticeably lower, compared to the baseline $k=1.0$. This suggests that if the true counterfactual association is substantially stronger than the calibrated lagged association, there might be more treatment effect heterogeneity than observed in the primary analysis. However, even in the $k=2.0$ scenario, the credible intervals still exhibit significant overlap across the strata.

Finally, Figure \ref{fig:PCE_delta_sensitivity} shows the sensitivity of stPCE ($d=1$) to the choice of the indifference threshold $\delta$, with $\delta$ varying from 0.5 to 1.5 (using the primary settings for $\rho_{\bm c}$ and $k=1.0$). The posterior estimates are robust to the definition of the principal strata, as the posterior summaries remain highly similar across different values of $\delta$.

\color{black}

\section{Discussion}\label{sec:discussion}
We have proposed an approach for inference on PCE with a continuous intermediate variable in SW-CRTs. Standard identifying assumptions include SUTVA and randomization for these longitudinal cluster randomized designs. In addition, we propose an identification assumption that uses a copula to identify the joint distribution of the intermediate variable. We also discuss a marginal structural assumption to link the unidentified conditional PCE with marginal means that can be identified from the observed data. We provide a way to calibrate the sensitivity parameters for both assumptions which exploits features of the SW-CRT design. We applied our method to analyze the HIV crowdsourcing study. {The results suggested a generally positive trend in HIV testing uptake due to the intervention across all principal strata. We found that the magnitude of the principal causal effects was highly similar across the different strata (dissociative and associative). This indicates that the intervention's effect on testing behavior does not significantly differ based on whether an individual's perceived social norms increased, decreased, or remained stable. The homogeneity suggests that the primary mechanism of the intervention's effect may not be strongly mediated by this specific measure of social norms. In short, our method provides methodology for inference on PCE in SW-CRT and an intuitive way to calibrate the sensitivity parameters.}

To the best of our knowledge, this is the first application of principal stratification framework to study the role of a continuous and repeatedly measured intermediate outcome in closed-cohort SW-CRTs.  Our proposed approach can be extended in several directions. {First, our approach assumes ignorable dropout to handle monotone attrition of individuals. However, non-ignorable dropout may occur especially in SW-CRTs for frail populations as patients with worsened outcomes may be more likely to discontinue the study. \citet{gasparini2024analysis} have introduced a joint modeling approach to address a certain type of non-ignorable dropout in closed-cohort SW-CRTs in the absence of intermediate outcomes, and it would be interesting to expand that approach in our setting. 
It is also worthwhile to consider pattern-mixture modeling frameworks that incorporate sensitivity parameters for handling missing data in longitudinal studies \cite{daniels2023bayesian}, and to develop analogous extensions for clustered longitudinal data structures that suit SW-CRTs. Second, in future work, we will also explore Bayesian nonparametric priors for the distribution of the observed data to further combat potential model misspecification bias for causal inference \citep{daniels2023bayesian}; the feasibility of such extensions have recently been demonstrated for cluster randomized trials with an intermediate outcome but in different contexts \citep{OHNISHI2025, YANG2025}.
Finally, if the interval for the sensitivity functions in Assumption \ref{assumption:msm} are too wide, an alternative prior would be to specify a double triangular prior which has its mode at the midpoint of the interval and decays to zero as move away from the mode (so basically, two triangular priors appropriately normalized). A formal implementation of such extensions is left for future work.}

\section*{Acknowledgments}
Daniels and Yang were partially supported by NIH R01 HL166324.  Li was supported by the Patient-Centered Outcomes Research Institute\textsuperscript{\textregistered} (PCORI\textsuperscript{\textregistered} Award ME-2023C1-31350). The statements presented in this article are solely the responsibility of the authors and do not necessarily represent the views of NIH, PCORI\textsuperscript{\textregistered}, its Board of Governors or Methodology Committee.

\section*{Data Availability Statement}
Access policies for the data set of the SW-CRT analyzed in our manuscript can be found in \citet{TANG2018} at \url{https://doi.org/10.1371/journal.pmed.1002645}. The analysis code is available at \href{https://github.com/lyang972/ps_swcrt/}{https://github.com/lyang972/ps\_swcrt/}.

\bibliographystyle{biom}  
\bibliography{references}

\end{document}